\documentclass{article}

\usepackage{arxiv}

\usepackage[utf8]{inputenc} 
\usepackage[T1]{fontenc}    
\usepackage{hyperref}       
\usepackage{url}            
\usepackage{booktabs}       
\usepackage{amsfonts}       
\usepackage{nicefrac}       
\usepackage{microtype}      
\usepackage{lipsum}		
\usepackage{graphicx}
\usepackage{natbib}
\usepackage{doi}
\usepackage{lscape}
\usepackage{subfigure}
\usepackage{minipage}

\title{Distinct polytropic behavior of plasma during ICME-HSS Interaction}

\author{ Kalpesh Ghag$^{1}$,
Anil Raghav$^{1}$\thanks{raghavanil1984@gmail.com},
Zubair Shaikh$^{2}$,
Georgios Nicolaou$^{3}$,
Omkar Dhamane$^{1}$,
Utsav Panchal$^{1}$\\
$^{1}$Department of Physics, University of Mumbai,Vidyanagari, Santacruz (E), Mumbai 400098, India\\
$^{2}$Indian Institute of Geomagnetism (IIG), New Panvel, Navi, Mumbai 410218, India\\
$^{3}$Department of Space and Climate Physics, Mullard Space Science Laboratory, \\
University College London, Dorking, Surrey, RH5 6NT, UK}

\renewcommand{\shorttitle}{\textit{arXiv} Template}

\hypersetup{
pdftitle={A template for the arxiv style},
pdfsubject={q-bio.NC, q-bio.QM},
pdfauthor={David S.~Hippocampus, Elias D.~Striatum},
pdfkeywords={First keyword, Second keyword, More},
}

\begin{document}
\maketitle

\begin{abstract}
	Interplanetary Coronal Mass Ejections (ICMEs) and High Speed Streams (HSSs) are noteworthy drivers of disturbance of interplanetary space. Interaction between them can cause several phenomena, such as; generation of waves, enhanced  geo-effectiveness, particle acceleration, etc. However, how does thermodynamic properties vary during the ICME-HSS interaction remain an open problem. In this study, we investigated the polytropic behavior of plasma during an ICME-HSS interaction observed by STEREO and Wind spacecraft. We find that the ICME observed by the STEREO-A has polytropic index  $\alpha = 1.0$, i.e., exhibiting isothermal process. Moreover, Wind spacecraft observed the HSS region, non-interacting ICME, and ICME-HSS interaction region. During each regions we found  $\alpha$=1.8, $\alpha$=0.7, and $\alpha$=2.5, respectively. It implies that the HSS region exhibits a nearly adiabatic behaviour, ICME region is closely isothermal, and the ICME-HSS interaction region exhibits super-adiabatic behaviour. The insufficient expansion of the ICME due to the interaction with HSS triggers the system for heating and cooling mechanisms which dependent on the degrees of freedom of plasma components.
\end{abstract}


\section{Introduction}
Thermodynamic studies include the investigation of relationship between heat, work, temperature, and energy. Studies of polytropic processes in plasmas, offer a novel way to investigate the plasma thermodynamics through the relationships between its macroscopic bulk parameters. Polytropic processes are quasi-stable processes in which the specific heat remains constant. For a plasma (an ideal gas case) it is defined as \citep{livadiotis2019origin,nicolaou2014long,nicolaou2020polytropic}; 
    \begin{equation}
        P_{th}    \propto  N ^{\alpha}
    \end{equation}
where, $P_{th}$ is thermal pressure $P_{th}$, $N$ is plasma density , and $\alpha$ is the polytropic index. $\alpha$ is also defined in terms of specific heats $\alpha = \frac{c_p - c}{c_v - c}$ (c = $\frac{dQ}{dT}$, Q is heat, T is temperature, and $c_p$ and $c_v$ are specific heat at constant pressure and volume respectively) \citep{Newbury1997}. We can determine the polytropic index through  correlations of the plasma moments and we can use it to study the plasma thermodynamics. The different values of $\alpha$ describe several thermodynamic processes of a system, e.g., an isothermal process ($\alpha = 1$), an isobaric process ($\alpha = 0$), an isochoric process ($\alpha \sim \infty$), and so on \citep{livadiotis2012non,nicolaou2014long}. 
Note that $\alpha$ is different from the well known quantity $\gamma = \frac{c_p}{c_v}$ i.e. $c_p$ and $c_v$ are specific heat at constant pressure and volume, respectively. The value of $\alpha$=$\gamma$ is found only in the spacial case (adiabatic case), in which there is no heat tranfrer (c= $\frac{dQ}{dT}$ =0). 
Different studies investigated the polytropic index for different astrophysical domains, e.g.  galaxy cluster ($\alpha \sim  1.2 - 1.3$) and galaxy supercluster  ($\alpha \sim 1.16$)  \citep{markevitch1998temperature,ettori2000bepposax}, solar corona ($\alpha = 1.10 - 1.58$) \citep{van2011first,prasad2018polytropic},  ($\alpha = 1.12$) of upper chromosphere \citet{houston2018magnetic}, hot flare loop $\alpha = 1.64$ \citep{wang2015evidence}, solar flare  $\alpha = 1.64 - 1.66$ \citep{Garcia2001, wang2015evidence}, solar wind  ($\alpha = 1.46 - 1.67$) \citep{totten1995empirical,Newbury1997,Livadiotis2016,nicolaou2019long}, bow-shock ($\alpha =$ 1.85) and $\alpha \sim$ 1.67 for earth's plasma sheet \citep{Tatrallyay1984,zhu1990plasma}, e.t.c. nterplanetary Coronal Mass Ejections (ICMEs) are observed to have a polytropic index ranging from 1.1 to 1.3, depending on where they are observed \citep{liu2006thermodynamic,osherovich1993polytropic}. Recently, \citet{mishra2018modeling} proposed that the polytropic index of a CME decreases from $\alpha =$ 1.87 close to the Sun, to $\alpha =$ 1.3 at 1 au i.e., $\alpha$ decreases as CME propagates in interplanetary space.   

ICMEs are magnetised plasma eruptions from the outer atmosphere of sun famously known as corona and observed in the heliosphere using in-situ measurements \citep{zurbuchen2006situ,kilpua2017coronal}.  These structures cause major disturbances in the heliosphere. The other structures which majorly contributes in the heliospheric disturbance are Stream Interaction Regions (SIRs) and Co-rotating Interaction Regions (CIRs). An ICME has three distinct parts i.e. forward propagating shock, sheath, and the orderly magnetised flux rope which remain magnetically connected to the Sun \citep{burlaga1988magnetic,zurbuchen2006situ}. Moreover, as an ICME propagates in the heliosphere it interacts with the surrounding solar wind and occasionally with SIRs and/or CIRs. Some times it interacts with the fast solar wind known as High Speed Stream (HSS). Such an interaction causes change in the ICME plasma speed, density, pressure, magnetic field, and its structure \citep{winslow2016longitudinal}. When ICME interacts with HSS it faces several major effects such as kink or deformation \citep{manchester2004modeling}, erosion due to magnetic reconnection \citep{dasso2006new},  deflection from its usual trajectory \citep{manchester2005coronal,wang2014deflected,wang2016propagation}, rotation through different axes, etc. \citet{nieves2012remote} used multi-spacecraft observations to study the signifficant reoriantaion of an ICME due to its interaction with solar wind. It is interesting to know the thermodynamic behavior of ICME in regimes with HSS interactions. 

It is important to note that the change in ICME plasma properties or structures when interacting with a HSS or SIR/CIR, depends on the whether ICME is hitted from the back or from the front, e.g., either the HSS or SIR/CIR overcome the ICME, or the ICME overcomes the HSS or SIR/CIR, respectively. Generally in ICME–HSS interaction events, researchers often focused on HSSs catching up a CME. Such interaction causes huge compression of the ICME from behind, hence deformation \citep{he2018stealth,heinemann2019cme}.  Some times, these interactions cause enhancement in turbulence within the ICME sheath region \citep{kilpua2017coronal}. The ICME-HSS, ICME-CIR, ICME-SIR, and ICME-ICME interaction causes several physical processes such as; charge particle acceleration \citep{richardson2003cme,zhuang2020role,lugaz2017interaction,gopalswamy2012factors,morosan2020electron}, generation of  Alf\'{e}ven waves \citep{raghav2018first,raghav2018torsional,raghav2018does,raghavdoes}, extremely intense geo-effectiveness \citep{shaikh2019concurrent,scolini2020cme,raghav2018torsional}, etc.

Recently, \citet{lugaz2022coronal} investigated an ICME observed on February 23 2021 by the  Solar TErrestrial RElations Observatory (STEREO)-A and Wind spacecrafts (separated longitudinally by $ 55 ^{o}$) using \textit{in-situ}  and remote-sensing measurements. They claimed that STEREO observed only ICME signatures whereas Wind spacecraft detected an ICME-HSS interaction. The \textit{in-situ} observations from STEREO-A and Wind allow us to examine the polytropic behavior of ICME plasma at different locations in space. Whereas the inhomogeneity in the interaction with HSS help us to probe the different thermodynamical process in a single ICME, e.g. a non-interacting ICME at STEREO-A and the HSS interacting with an ICME at Wind. This observation gives us a great opportunity to study such a dynamic interaction and its consequences to the plasma thermodynamics, using the polytropic process approach.

\section{Data and Methods}

We have used \textit{in situ} data from STEREO-A and Wind spacecraft to study interplanetary parameters and the polytropic behaviour of an ICME-HSS interaction.  The Solar Wind Experiment (SWE; \cite{ogilvie1995swe}) instrument on board the Wind spacecraft measures plasma bulk properties, such as proton speed $V_p$, proton number density $N_p$, and proton temperature $T_p$, while the Magnetic Field Investigation (MFI; \cite{lepping1995wind}) instrument measures the magnetic field. In STEREO PLasma And Supra-Thermal Ion Composition (PLASTIC) \citep{galvin2008plasma} instrument gives solar wind parameter where the magnetometer gives the magnetic field components similar to Wind. For polytropic index analysis we used thermal pressure($P_{th} = N_p k_B T_p$) and $N_p$. By taking the natural logarithm of Eq. 1, we get
\begin{equation}
     \ln{P_{th}} = \alpha \ln{N_p} + \ln{F} .
    \end{equation}
    
    We then perform a linear fitting to $\ln{P_{th}}$ vs $\ln{N_p}$. The slope of the fitted line, is the polytropic index $\alpha$. The y-intersection gives the constant of the equation, i.e.,$\ln{F}$. First, we calculated the polytropic index of the magnetic cloud as observed by STEREO-A. We then used the Wind observations to determine a for the following regions : i) HSS observed on 20 - 23 February 2021, ii) Magnetic cloud observed on 24 February 2021, and iii) Magnetic cloud region merged in HSS at the later half of 24 February, 2021, respectively (See figure 5 of \cite{lugaz2022coronal}).

\section{Observations and Results}
We have performed detail analysis of ICME-HSS interaction event discussed by the \cite{lugaz2022coronal} using multi-spacecraft data. The multi-spacecraft observation of ICME is confirmed by the catalogue given by \cite{mostl2022multipoint}. The boundaries of different regions are taken from the above articles.  
\subsection{STEREO-A }

Figure \ref{IP} (a) shows the plasma parameters observed at STEREO-A during ICME transit on  23-24 February, 2021. The sudden rise in Interplanetary Magnetic Field (IMF), proton velocity ($V_p$) and proton number density ($N_p$) shows signature of shock (first vertical dash line) on 23$^{rd}$ February at 10.34 UT. The sheath region (yellow shaded region) is observed from shock onset to the start of magnetic cloud at 01.00 UT on 24$^{th}$ February.  The low plasma beta, low proton number density, and the rotation in IMF vectors indicates the signature of magnetic cloud (MC; see the green shaded region) starting at 01.00 UT on 24$^{th}$ February and ends on 24$^{th}$ February at 23:00 UT. During MC transit the magnetic field was steady and almost entirely radial \citep{lugaz2022coronal}. Figure 1(d) shows variations of ln(Pth) as a function of ln(Np) for the MC interval. In the same panel we show the linear fitting of Eq. 2 to the data. We observed that the $\alpha$= 1.0 during MC transit. It implies that MC observed by the STEREO-A shows isothermal property.

\subsection{Wind }
\begin{landscape}
\begin{figure}
\centering
\vspace{-1.9cm}
\hspace{-1.8cm}
\begin{minipage}[t]{7.6cm}
\centering
\textbf{(a)  STEREO A (ICME)}\par
\subfigure{\includegraphics[scale=0.43]{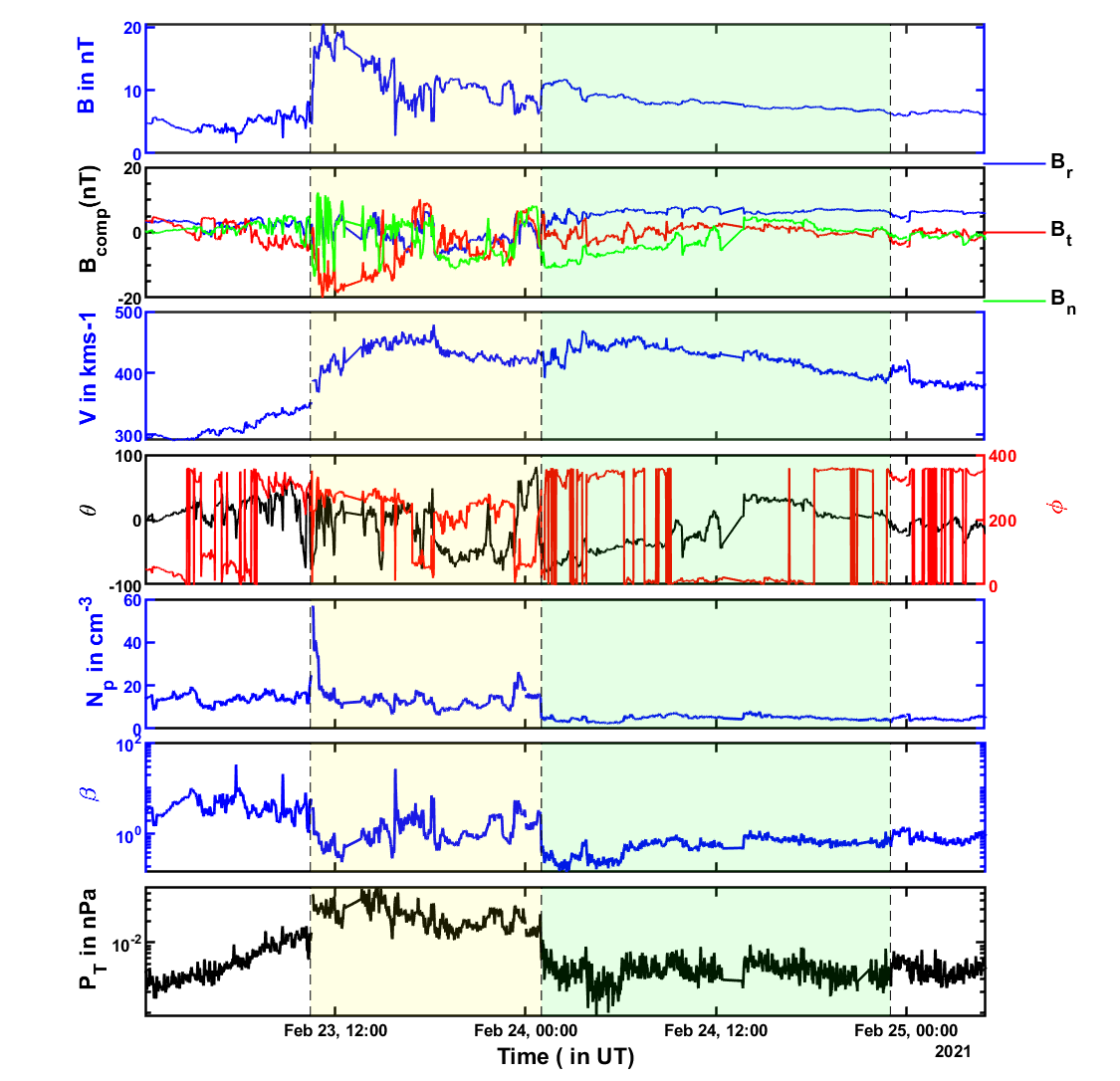}}
\end{minipage}
\hspace{0.48cm}
\begin{minipage}[t]{7.55cm}
\centering

\textbf{(b) Wind (HSS)}\par
\subfigure{\includegraphics[scale=0.43]{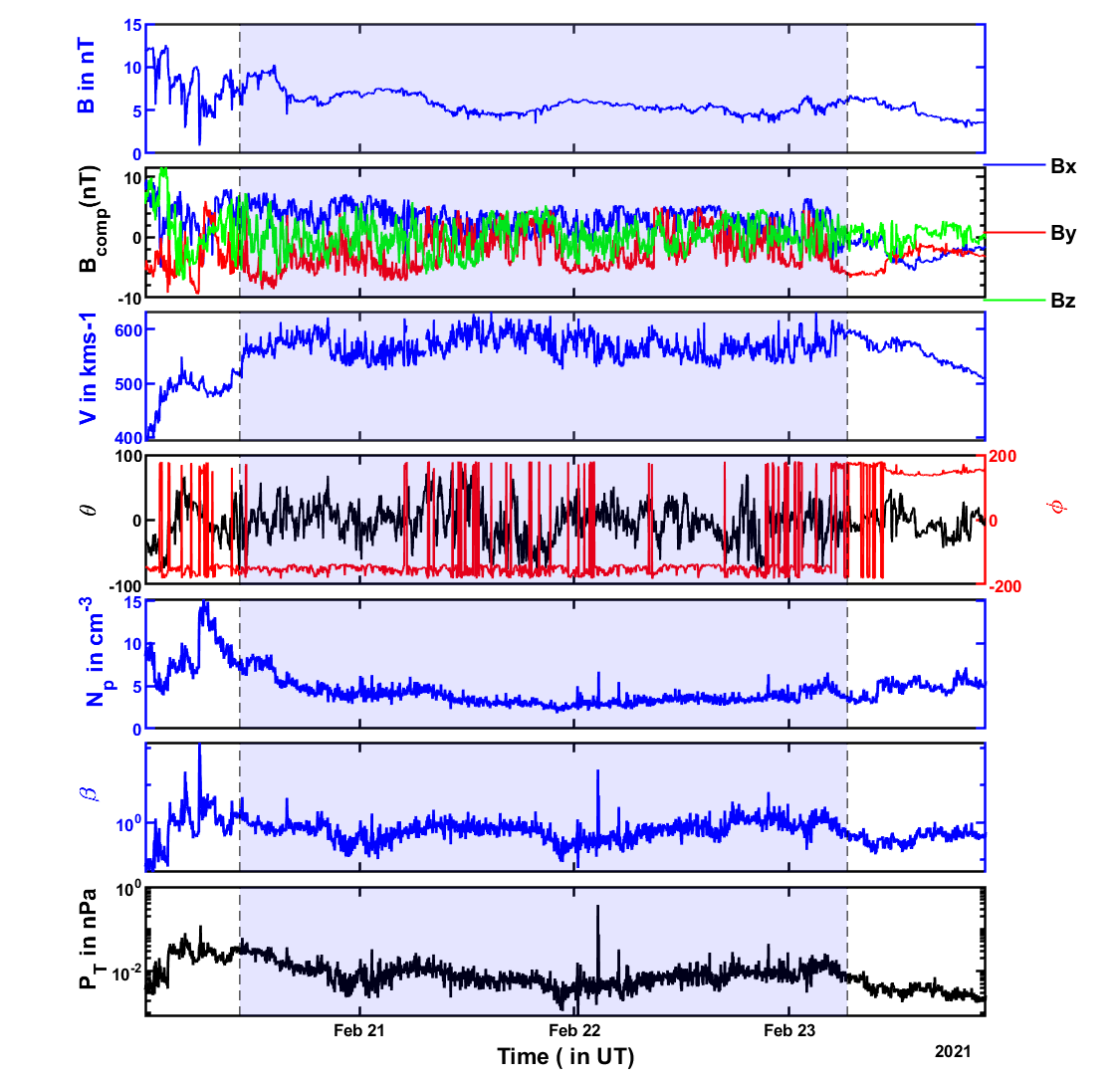}}
\end{minipage}
\hspace{0.65cm}
\begin{minipage}[t]{7.5cm}
\centering
\textbf{(c) Wind (ICME)}\par
\subfigure{\includegraphics[scale=0.43]{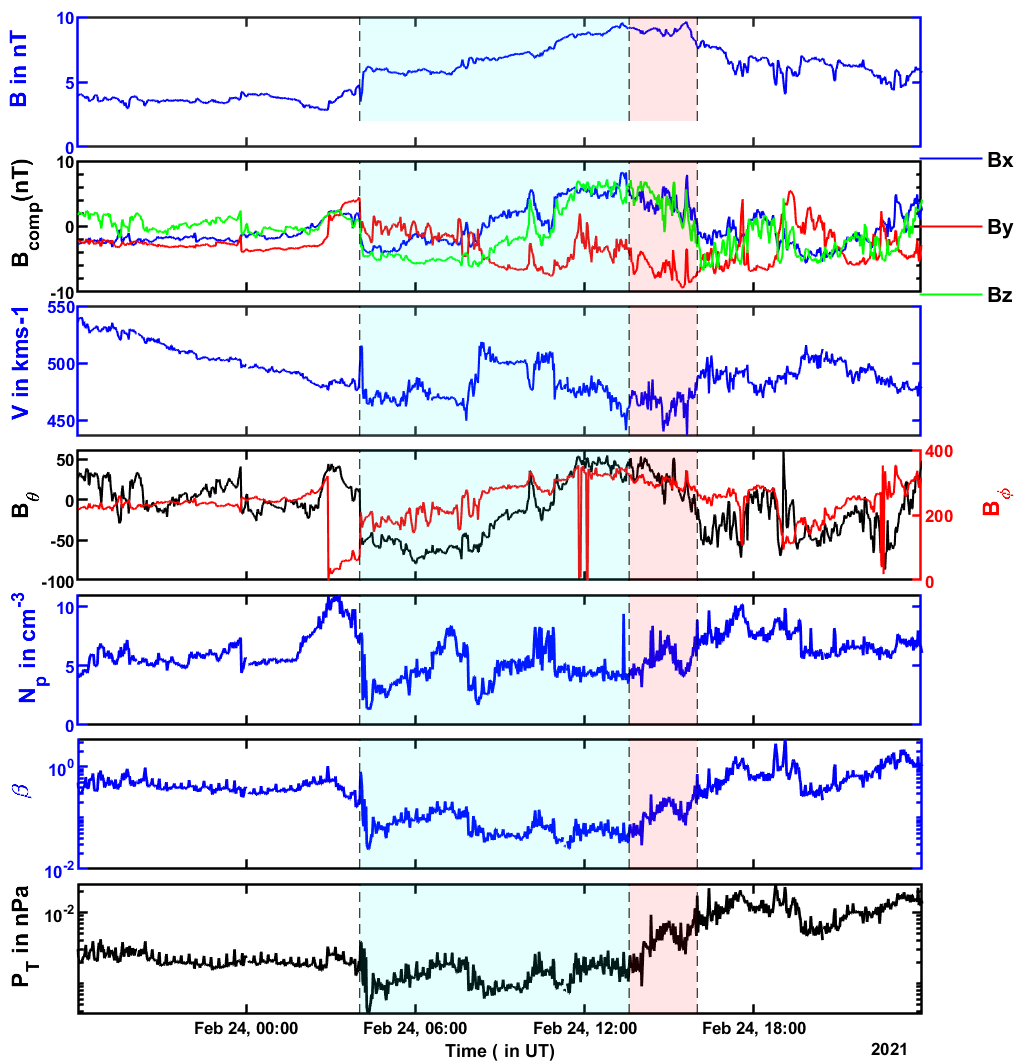}}
\end{minipage}

\hspace{-1.8cm}
\begin{minipage}[t]{7.5cm}
\centering
\textbf{(d)  }
\subfigure{\includegraphics[scale=0.3]{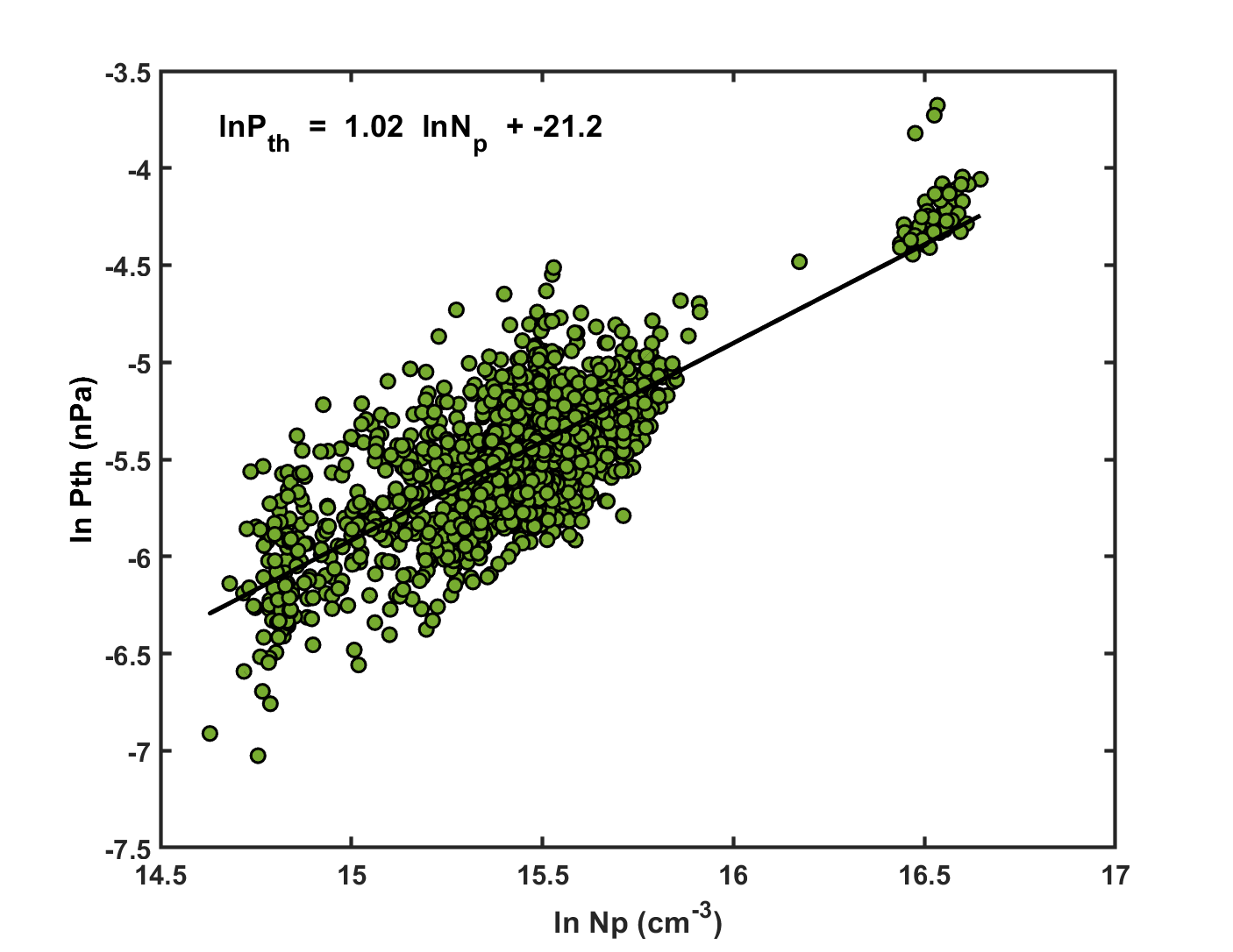}}
\end{minipage}
\hspace{0.38cm}
\begin{minipage}[t]{7.5cm}
\centering
\textbf{(e)  }
\subfigure{\includegraphics[scale=0.3]{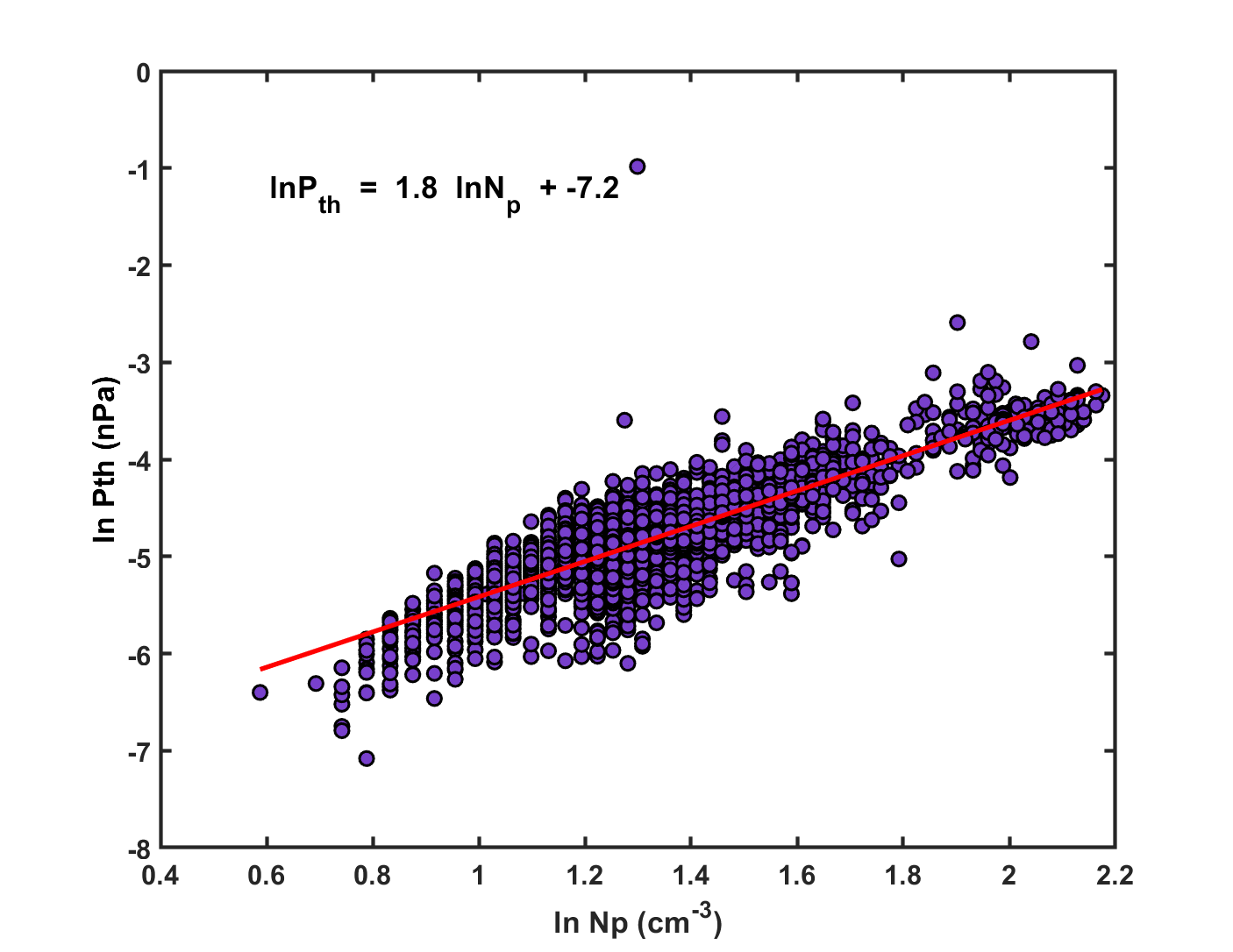}}
\end{minipage}
\hspace{0.38cm}
\begin{minipage}[t]{7.5cm}
\centering
\textbf{(f) }
\subfigure{\includegraphics[scale=0.3]{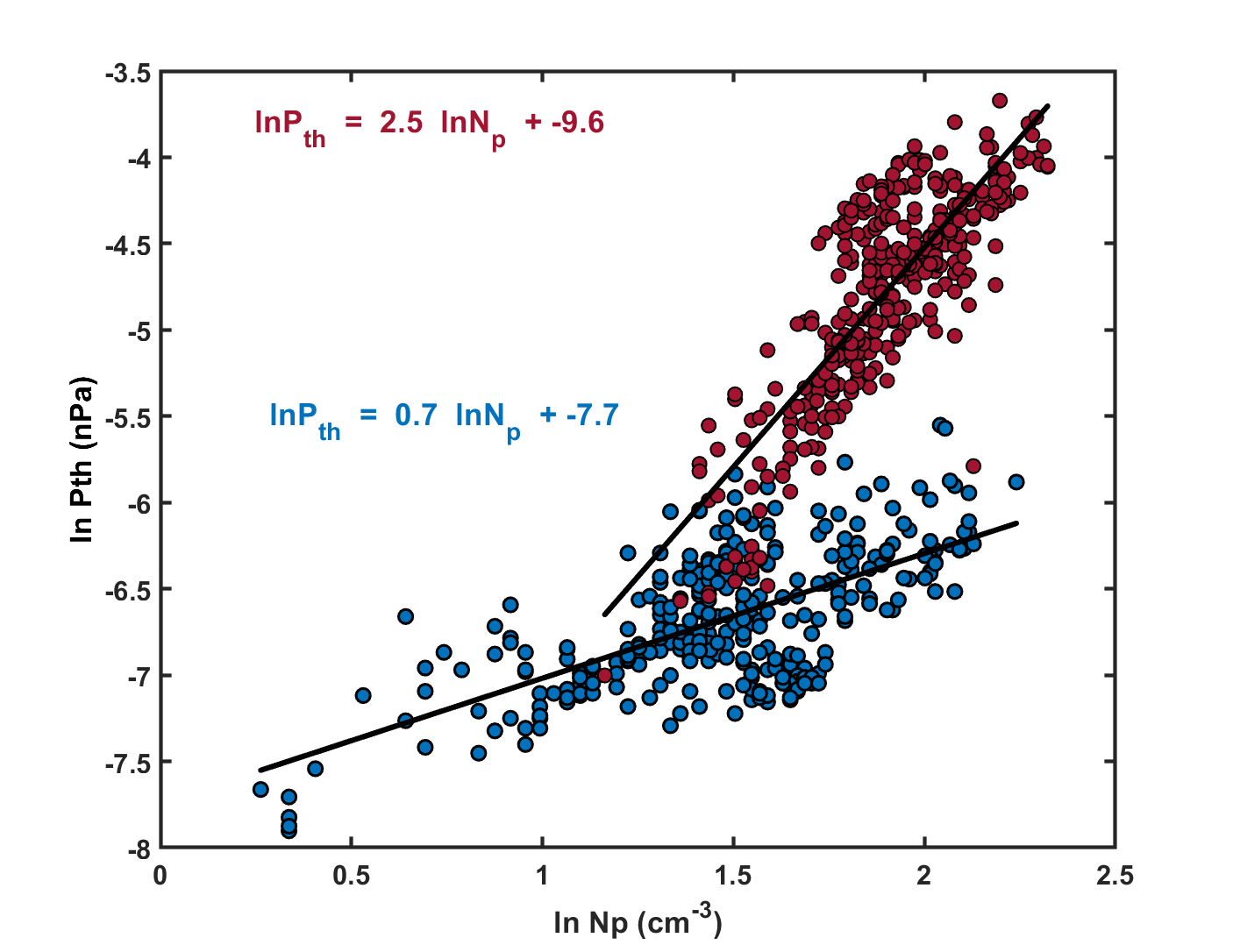}}
\end{minipage}
\caption{\textit{ The observation of : (a) STA (ICME),  (b) Wind (HSS), and (c) Wind (ICME) spacecrafts from February 20-24, 2021. From top panel shows the i) total magnetic field ii) Magnetic field component ($B_x$, $B_y$, $B_z$) iii) Proton velocity($V_p$) iv) The inclination angle of magnetic field vector ($\theta$) and azimuthal angle ($\phi$)  v) Proton number density ($N_p$), vi) Plasma beta ($\beta$) vii) Thermal pressure ($P_{th}$). For figure (a) yellow shaded portion shows sheath and green shaded portion shows magnetic cloud (MC). For figure (b) purple shaded region shows HSS. In figure (c) cyan shaded region shows non interacting magnetic cloud where red shaded region shows the ICME-HSS interaction. The second row represents polytropic analysis i.e. variation of $\ln{P_{th}}$ as a function of $\ln{N_{p}}$ the of above events observed at STEREO and Wind. For figure (d) green colored circles represents the data points for magnetic cloud of ICME observed by STEREO-A. In figure (e), purple colored circle shows data points for HSS region observed by Wind. In figure (f), cyan coloured circles represents data points of pure magnetic cloud and red coloured circles represents data points of magnetic cloud which is in interaction with HSS. The solid lines give linear fitting values.}}
\label{IP}
\end{figure}
\end{landscape}

The plasma parameters during the HSS are shown by Figure \ref{IP} (b). The enhanced proton velocity shows the signatures of HSS. Figure \ref{IP} (c) displays the plasma parameters and magnetic field variation of the MC observed by the Wind spacecraft on 24th February 2021. The low plasma beta, low number density and the smooth rotation of the magnetic field signifies the magnetic cloud. This is the same ICME MC observed by the STEREO-A spacecraft. The MC interacted at the trailing part of HSS which was observed from 20 February 2021, 12.00 UT to  23 February, 23.00 UT, 2021 \citep{lugaz2022coronal}. The ICME MC starts on 24 February, 2021 04.00 UT and increase in proton number density and proton velocity suggest us to denote the end time as 16.00 UT on 24 February, 2021.  The ICME does not show any shock and sheath signature. Furthermore, we have calculated the $\alpha$ during the following region (Fig. (e) and (f) in Fig. \ref{IP}): i) HSS from 20 - 23 February, 2021; ii) MC from 04:00 UT to 16:00 UT on 24 February 2021; iii) MC merged in HSS on 24 February, 2021.  We found  $\alpha = 1.8$, $\alpha = 0.7$, and $\alpha = 2.5$ for the above regions respectively. Thus, different $\alpha$ value is observed at different regions during ICME-HSS interaction.

\section{Discussion}

This study focuses on the polytropic behavior of ICME and its interaction with high speed ambient solar wind.  In previous studies the polytropic index of solar wind proton is empirically calculated as 1.46  \citep{totten1995empirical}.  Furthermore, \cite{Newbury1997} derived the polytropic index in the vicinity of stream interactions. The value asymptotically found around 1.67 (i.e. 5/3) and in some occasions  $\alpha \sim $ 2. \cite{nicolaou2014long} studied the long term variation of polytropic index over 17 years, where they found the the $\alpha \sim$ 1.8 on average. Later, \cite{nicolaou2020polytropic} had verified these results using the high resolution measurements by Parker Solar Probe. The large-scale variations of the solar wind proton density and temperature, which are associated with the plasma expansion into the heliosphere, follow a polytropic model with a polytropic index $5/3$.  Furthermore, many studies have observed the polytropic index of the fast and slow solar wind deviates upto 1.8 to 2.0. It is interesting to note that all these studies found that the polytropic behaviour of solar wind plasma is independent on the solar wind speed \citep{Newbury1997,nicolaou2019long,livadiotis2018thermodynamic}. We found the $\alpha$  = 1.8 for the HSS region interpreted as near adiabatic behaviour.  Besides the significant variations of the proton bulk parameters, $\alpha$ is nearly constant. 

For ICME, \cite{osherovich1993polytropic} noted different polytropic index within the MC for proton ($\alpha = 1.2$). It suggest that protons within the MC has quasi-isothermal characteristic. \cite{liu2006thermodynamic} also show that the ICME has isothermal properties. It implies that, the heat flows from system under constant temperature. Recently, \cite{mishra2018modeling} found that the $\alpha$ for CME varies from $\alpha =$ 1.87 to $\alpha =$ 1.3 as it propagates away from the sun. Initially, when CME emerges from the vicinity of sun, its temperature is high with respect to ambient solar wind. During the erruption of a CME, it is expected as it is being pushed from the back by the high-speed wind, which contributed to the CME’s propagation speed but prohibited the CME from expansion. The insufficient expansion restrict the $\alpha$ > 1.67.

As it propagates through heliosphere, the expansion of CME causes decrease in temperature as well as number density. The CME reaches to the adiabatic case where the heat is not flowing from system to surrounding \citep{mishra2018modeling}. Later $\alpha$ starts to decrease via transferring heat to the surrounding plasma and reaches to quasi-isothermal state. It implies that ICME close to the Sun has adiabatic characteristic whereas at 1 AU it shows quasi-isothermal properties. In our study, we have found $\alpha$ = 1.0 at the one arm of CME observed by the STEREO-A, which is in good agreement with \cite{liu2006thermodynamic} and \cite{osherovich1993polytropic}. On the other hand, ICME MC observed by the Wind spacecraft at 1 AU found $\alpha = 0.70$. Note that, the time gap between HSS and ICME is larger at STEREO-A hence our results are not affected by any interaction. Whereas, at the Wind location the separation between HSS and ICME was small and therefore, the slighlty lower $\alpha$ we observe might be a result of the interaction between the two structures.

The ICME interacts with HSS from the side which is crossed by Wind spacecraft on the second half of 24 February, 2022. We have estimated the polytropic index ($\alpha$) for this region, surprisingly, we found the $\alpha = 2.54$ during ICME-HSS interaction region. This suggests that investigated region shows super-adiabatic characteristic \citep{livadiotis2019origin,nicolaou2014long,nicolaou2020polytropic}. We believe that such high value of $\alpha$ during ICME-HSS interaction could be due to high compression which changes the plasma properties at the time of interaction. The ICME-HSS interaction can restrict the magnetic cloud and solar wind to expand. The insufficient expansion might not have allowed the CME to be cool enough to depart from the heat releasing state i.e isothermal to an superadiabatic  state. The MC changes its state from ($\alpha$)= 1.0 to ($\alpha$)= 2.54 during interaction. Due to the interaction with the HSS, the restricted expansion of MC implies the work on the system. Note that the heating and cooling process of plasma depends on the kinetic degrees of freedom of the system.

Therefore, to quantify the heating and cooling process, we have used the relationship between $\gamma$ and degrees of freedom ($f$) as $ f = \frac{2}{\gamma- 1}$. From the first law of thermodynamics and following algebra by \cite{livadiotis2019connection}, we get
\begin{equation}
      \alpha = \frac{2}{f} \Bigg[ 1 - \frac{\delta q}{\delta w} \Bigg] + 1
\end{equation}

\begin{figure}
    \centering
    \includegraphics[scale = 0.5]{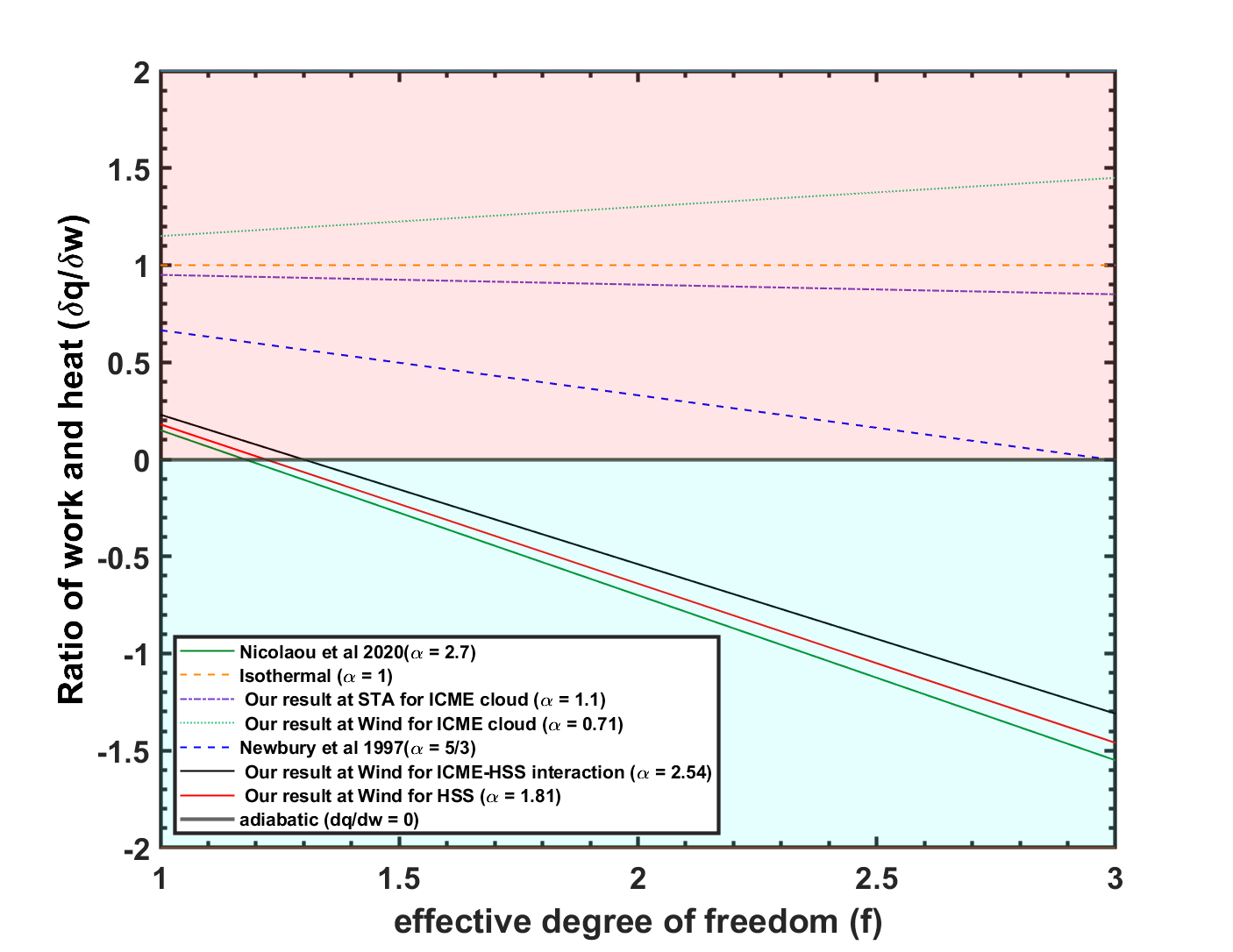}
    \caption{The relationship between the $\frac{\delta q}{\delta w}$ and $f$ for various $\alpha$ values. The heating and cooling systems are represented by the red- and cyan-shaded areas. The various lines shows the $\frac{\delta q}{\delta w}$ variation over $f$ for different results}
    \label{Cooling-heating}
\end{figure}

Figure \ref{Cooling-heating} shows the  relationship between the $\frac{\delta q}{\delta w}$ and $f$ for various $\alpha$ values. The red shaded region shows the heating process i.e. $\frac{\delta q}{\delta w}$ > 0 whereas the cyan shaded region shows the cooling processes in expanding plasma for  $\frac{\delta q}{\delta w}$ < 0. For isothermal process where $\alpha$ = 1, the system is always under the circumstances of heating for any value of $f$.  For $\alpha$ = $\gamma$ = 5/3 the system is under heating process for the $f$ < 3. At $f$= 3, the system is adiabatic. \cite{nicolaou2020polytropic} has measured the $\alpha$ value using PSP data between 0.17 AU to 0.80 AU. It is found that the alpha value is about 2.7 near the vicinity of sun. From Figure \ref{Cooling-heating} it is observed that for $f$ = 1.2 the process is adiabatic for the given polytropic index. This means the heat exchange between system and surrounding is zero. For $f$ < 1.2 the $\frac{\delta q}{\delta w}$ is positive which implies that the the heat is gained by the system and hence the heating process may be dominant. On contrast for $f$ > 1.2 the negative value of $\frac{\delta q}{\delta w}$ show the system looses its heat and cooling process may dominant. In our study, the magnetic clouds at both observation point i.e at STEREO-A and Wind, the process is isothermal. It implies that for any value of $f$ the heating process is dominant. Whereas, the a of the HSS region suggests that there is an energy exchange only if the plasma has $f$ <2.5, resulting to heating of the expanding plasma. The interaction region shows the $\frac{\delta q}{\delta w}$ is positive for the $f$ < 1.29. It implies that the heating of plasma occurs at $f$< 1.29. 

\section{Conclusion}

Our study examines the distinct polytropic behavior of plasma during the ICME- HSS interaction, here we conclude that.

\begin{enumerate}

    \item Th isolated ICME MC observed by the STEREO-A shows $\alpha = 1.0$, while MC observed by the Wind spacecraft shows $\alpha = 0.70$ suggesting that ICME MC shows nearly isothermal behavior.
    
    \item HSS has $\alpha \sim$ 1.8, which is nearly adiabatic behavior (for $f$=3)
    
    \item The interaction of ICME with HSS leads to magnetic cloud to behaving superadiabatic (for $f$ = 3 ) with $\alpha$ = 2.54.
    
    \item The insufficient expansion of the ICME due to the interacting HSS enables the system for heating and cooling mechanisms. Here the degrees of freedom plays vital role. For $f$ < 1.29 validates the kinetic description of plasma ions interacting with slow waves, where ions behave as if they are a one-dimensional ( $f$ = 1) adiabatic fluid ($\frac{\delta q}{\delta w}$ = 0) with temperature variations confined along the magnetic field \citep{Verscharen_2017, nicolaou2020polytropic}.
    
    \item The interaction might causes the alteration of magnetic field. The effective degrees of freedom are decreased when high magnetic fields dominate the thermal motions of the particles. Since, the thermodynamic processes are constrained along the magnetic field's direction and hence the system may efficiently absorbs energy. Thus, we believe that interaction of ICME with HSS causes significant change of thermodynamics of plasma.
    
    \item Our study gives the insight about the unusual behaviour of astrophysical space plasma. The macroscopic study using polytropic approach enables to enhance our understanding regarding the heat exchanges mechanism between different plasma structures and may form a corner stone in understanding the heating cooling process in more advance manner.
\end{enumerate}

\section*{Acknowledgements}
We acknowledge use of NASA/GSFC's Space Physics Data Facility's  CDAWeb service. We are thankful to DST, India, since KG is funded by DST-INSPIRE Fellowship (INSPIRE Fellow Registration Number: IF210212). We acknowledge SERB, India, since AR and OD is supported by SERB project reference file number CRG/2020/002314.

\section*{DATA AVAILABILITY}
The data in this analysis is taken from Wind and STEREO-A spacecrafts. The data is publicly available at (1) NASA's Goddard Space Flight Center (GSFC) \url{https://wind.nasa.gov/data.php}, and (2) Coordinated Data Analysis Web (CDAWeb)  \url{https://cdaweb.gsfc.nasa.gov/pub/data/wind/}.

\bibliographystyle{mnras}
\bibliography{Superadiabatic}  

\begin{thebibliography}{}
\makeatletter
\relax
\def\mn@urlcharsother{\let\do\@makeother \do\$\do\&\do\#\do\^\do\_\do\%\do\~}
\def\mn@doi{\begingroup\mn@urlcharsother \@ifnextchar [ {\mn@doi@}
  {\mn@doi@[]}}
\def\mn@doi@[#1]#2{\def\@tempa{#1}\ifx\@tempa\@empty \href
  {http://dx.doi.org/#2} {doi:#2}\else \href {http://dx.doi.org/#2} {#1}\fi
  \endgroup}
\def\mn@eprint#1#2{\mn@eprint@#1:#2::\@nil}
\def\mn@eprint@arXiv#1{\href {http://arxiv.org/abs/#1} {{\tt arXiv:#1}}}
\def\mn@eprint@dblp#1{\href {http://dblp.uni-trier.de/rec/bibtex/#1.xml}
  {dblp:#1}}
\def\mn@eprint@#1:#2:#3:#4\@nil{\def\@tempa {#1}\def\@tempb {#2}\def\@tempc
  {#3}\ifx \@tempc \@empty \let \@tempc \@tempb \let \@tempb \@tempa \fi \ifx
  \@tempb \@empty \def\@tempb {arXiv}\fi \@ifundefined
  {mn@eprint@\@tempb}{\@tempb:\@tempc}{\expandafter \expandafter \csname
  mn@eprint@\@tempb\endcsname \expandafter{\@tempc}}}

\bibitem[\protect\citeauthoryear{Burlaga}{Burlaga}{1988}]{burlaga1988magnetic}
Burlaga L.,  1988, Journal of Geophysical Research: Space Physics, 93, 7217

\bibitem[\protect\citeauthoryear{Dasso, Mandrini, D{\'e}moulin  \& Luoni}{Dasso
  et~al.}{2006}]{dasso2006new}
Dasso S.,  Mandrini C.~H.,  D{\'e}moulin P.,   Luoni M.~L.,  2006, Astronomy \&
  Astrophysics, 455, 349

\bibitem[\protect\citeauthoryear{Ettori, Bardelli, De~Grandi, Molendi, Zamorani
   \& Zucca}{Ettori et~al.}{2000}]{ettori2000bepposax}
Ettori S.,  Bardelli S.,  De~Grandi S.,  Molendi S.,  Zamorani G.,   Zucca E.,
  2000, Monthly Notices of the Royal Astronomical Society, 318, 239

\bibitem[\protect\citeauthoryear{Galvin et~al.,}{Galvin
  et~al.}{2008}]{galvin2008plasma}
Galvin A.~B.,  et~al., 2008, Space Science Reviews, 136, 437

\bibitem[\protect\citeauthoryear{Garcia}{Garcia}{2001}]{Garcia2001}
Garcia H.,  2001, \mn@doi [ApJ] {10.1086/321693}, 557, 897

\bibitem[\protect\citeauthoryear{Gopalswamy}{Gopalswamy}{2012}]{gopalswamy2012factors}
Gopalswamy N.,  2012, in AIP Conference Proceedings. pp 247--252

\bibitem[\protect\citeauthoryear{He, Liu, Hu, Wang  \& Zhao}{He
  et~al.}{2018}]{he2018stealth}
He W.,  Liu Y.~D.,  Hu H.,  Wang R.,   Zhao X.,  2018, The Astrophysical
  Journal, 860, 78

\bibitem[\protect\citeauthoryear{Heinemann et~al.,}{Heinemann
  et~al.}{2019}]{heinemann2019cme}
Heinemann S.~G.,  et~al., 2019, Solar Physics, 294, 1

\bibitem[\protect\citeauthoryear{Houston, Jess, Ramos, Grant, Beck, Norton  \&
  Prasad}{Houston et~al.}{2018}]{houston2018magnetic}
Houston S.,  Jess D.,  Ramos A.~A.,  Grant S.,  Beck C.,  Norton A.,   Prasad
  S.~K.,  2018, ApJ, 860, 28

\bibitem[\protect\citeauthoryear{Kilpua, Koskinen  \& Pulkkinen}{Kilpua
  et~al.}{2017}]{kilpua2017coronal}
Kilpua E.,  Koskinen H.~E.,   Pulkkinen T.~I.,  2017, Living Reviews in Solar
  Physics, 14, 1

\bibitem[\protect\citeauthoryear{Lepping et~al.,}{Lepping
  et~al.}{1995}]{lepping1995wind}
Lepping R.,  et~al., 1995, Space Science Reviews, 71, 207

\bibitem[\protect\citeauthoryear{Liu, Richardson, Belcher, Kasper  \&
  Elliott}{Liu et~al.}{2006}]{liu2006thermodynamic}
Liu Y.,  Richardson J.,  Belcher J.,  Kasper J.,   Elliott H.,  2006, Journal
  of Geophysical Research: Space Physics, 111

\bibitem[\protect\citeauthoryear{Livadiotis}{Livadiotis}{2018}]{livadiotis2018thermodynamic}
Livadiotis G.,  2018, EPL (Europhysics Letters), 122, 50001

\bibitem[\protect\citeauthoryear{Livadiotis}{Livadiotis}{2019a}]{livadiotis2019connection}
Livadiotis G.,  2019a, Entropy, 21, 1041

\bibitem[\protect\citeauthoryear{Livadiotis}{Livadiotis}{2019b}]{livadiotis2019origin}
Livadiotis G.,  2019b, ApJ, 874, 10

\bibitem[\protect\citeauthoryear{Livadiotis \& Desai}{Livadiotis \&
  Desai}{2016}]{Livadiotis2016}
Livadiotis G.,  Desai M.~I.,  2016, \mn@doi [ApJ] {10.3847/0004-637x/829/2/88},
  829, 88

\bibitem[\protect\citeauthoryear{Livadiotis \& McComas}{Livadiotis \&
  McComas}{2012}]{livadiotis2012non}
Livadiotis G.,  McComas D.,  2012, The Astrophysical Journal, 749, 11

\bibitem[\protect\citeauthoryear{Lugaz, Temmer, Wang  \& Farrugia}{Lugaz
  et~al.}{2017}]{lugaz2017interaction}
Lugaz N.,  Temmer M.,  Wang Y.,   Farrugia C.~J.,  2017, Solar Physics, 292, 1

\bibitem[\protect\citeauthoryear{Lugaz et~al.,}{Lugaz
  et~al.}{2022}]{lugaz2022coronal}
Lugaz N.,  et~al., 2022, The Astrophysical Journal, 929, 149

\bibitem[\protect\citeauthoryear{Manchester~IV, Gombosi, Roussev, Ridley,
  De~Zeeuw, Sokolov, Powell  \& T{\'o}th}{Manchester~IV
  et~al.}{2004}]{manchester2004modeling}
Manchester~IV W.~B.,  Gombosi T.~I.,  Roussev I.,  Ridley A.,  De~Zeeuw D.~L.,
  Sokolov I.,  Powell K.~G.,   T{\'o}th G.,  2004, Journal of Geophysical
  Research: Space Physics, 109

\bibitem[\protect\citeauthoryear{Manchester~IV et~al.,}{Manchester~IV
  et~al.}{2005}]{manchester2005coronal}
Manchester~IV W.,  et~al., 2005, The Astrophysical Journal, 622, 1225

\bibitem[\protect\citeauthoryear{Markevitch, Forman, Sarazin  \&
  Vikhlinin}{Markevitch et~al.}{1998}]{markevitch1998temperature}
Markevitch M.,  Forman W.~R.,  Sarazin C.~L.,   Vikhlinin A.,  1998, The
  Astrophysical Journal, 503, 77

\bibitem[\protect\citeauthoryear{Mishra \& Wang}{Mishra \&
  Wang}{2018}]{mishra2018modeling}
Mishra W.,  Wang Y.,  2018, ApJ, 865, 50

\bibitem[\protect\citeauthoryear{Morosan et~al.,}{Morosan
  et~al.}{2020}]{morosan2020electron}
Morosan D.,  et~al., 2020, Astronomy \& Astrophysics, 642, A151

\bibitem[\protect\citeauthoryear{M{\"o}stl et~al.,}{M{\"o}stl
  et~al.}{2022}]{mostl2022multipoint}
M{\"o}stl C.,  et~al., 2022, The Astrophysical Journal Letters, 924, L6

\bibitem[\protect\citeauthoryear{Newbury, Russell  \& Lindsay}{Newbury
  et~al.}{1997}]{Newbury1997}
Newbury J.~A.,  Russell C.~T.,   Lindsay G.~M.,  1997, \mn@doi [GRL]
  {10.1029/97gl01204}, 24, 1431

\bibitem[\protect\citeauthoryear{Nicolaou \& Livadiotis}{Nicolaou \&
  Livadiotis}{2019}]{nicolaou2019long}
Nicolaou G.,  Livadiotis G.,  2019, The Astrophysical Journal, 884, 52

\bibitem[\protect\citeauthoryear{Nicolaou, Livadiotis  \& Moussas}{Nicolaou
  et~al.}{2014}]{nicolaou2014long}
Nicolaou G.,  Livadiotis G.,   Moussas X.,  2014, Solar Physics, 289, 1371

\bibitem[\protect\citeauthoryear{Nicolaou, Livadiotis, Wicks, Verscharen  \&
  Maruca}{Nicolaou et~al.}{2020}]{nicolaou2020polytropic}
Nicolaou G.,  Livadiotis G.,  Wicks R.~T.,  Verscharen D.,   Maruca B.~A.,
  2020, ApJ, 901, 26

\bibitem[\protect\citeauthoryear{Nieves-Chinchilla, Colaninno, Vourlidas,
  Szabo, Lepping, Boardsen, Anderson  \& Korth}{Nieves-Chinchilla
  et~al.}{2012}]{nieves2012remote}
Nieves-Chinchilla T.,  Colaninno R.,  Vourlidas A.,  Szabo A.,  Lepping R.,
  Boardsen S.,  Anderson B.,   Korth H.,  2012, Journal of Geophysical
  Research: Space Physics, 117

\bibitem[\protect\citeauthoryear{Ogilvie et~al.,}{Ogilvie
  et~al.}{1995}]{ogilvie1995swe}
Ogilvie K.,  et~al., 1995, Space Science Reviews, 71, 55

\bibitem[\protect\citeauthoryear{Osherovich, Farrugia, Burlaga, Lepping,
  Fainberg  \& Stone}{Osherovich et~al.}{1993}]{osherovich1993polytropic}
Osherovich V.,  Farrugia C.,  Burlaga L.,  Lepping R.,  Fainberg J.,   Stone
  R.,  1993, JGR: Space Physics, 98, 15331

\bibitem[\protect\citeauthoryear{Prasad, Raes, Van~Doorsselaere, Magyar  \&
  Jess}{Prasad et~al.}{2018}]{prasad2018polytropic}
Prasad S.~K.,  Raes J.,  Van~Doorsselaere T.,  Magyar N.,   Jess D.,  2018,
  ApJ, 868, 149

\bibitem[\protect\citeauthoryear{Raghav \& Kule}{Raghav \& Kule}{}]{raghavdoes}
Raghav A.~N.,  Kule A., , Monthly Notices of the Royal Astronomical Society:
  Letters

\bibitem[\protect\citeauthoryear{Raghav \& Kule}{Raghav \&
  Kule}{2018a}]{raghav2018first}
Raghav A.~N.,  Kule A.,  2018a, MNRAS: Letters, 476, L6

\bibitem[\protect\citeauthoryear{Raghav \& Kule}{Raghav \&
  Kule}{2018b}]{raghav2018does}
Raghav A.~N.,  Kule A.,  2018b, Monthly Notices of the Royal Astronomical
  Society: Letters, 480, L6

\bibitem[\protect\citeauthoryear{Raghav, Kule, Bhaskar, Mishra, Vichare  \&
  Surve}{Raghav et~al.}{2018}]{raghav2018torsional}
Raghav A.~N.,  Kule A.,  Bhaskar A.,  Mishra W.,  Vichare G.,   Surve S.,
  2018, ApJ, 860, 26

\bibitem[\protect\citeauthoryear{Richardson, Lawrence, Haggerty, Kucera  \&
  Szabo}{Richardson et~al.}{2003}]{richardson2003cme}
Richardson I.~G.,  Lawrence G.~R.,  Haggerty D.~K.,  Kucera T.~A.,   Szabo A.,
  2003, Geophysical research letters, 30

\bibitem[\protect\citeauthoryear{Scolini et~al.,}{Scolini
  et~al.}{2020}]{scolini2020cme}
Scolini C.,  et~al., 2020, The Astrophysical Journal Supplement Series, 247, 21

\bibitem[\protect\citeauthoryear{Shaikh, Raghav, Vichare, Bhaskar, Mishra  \&
  Choraghe}{Shaikh et~al.}{2019}]{shaikh2019concurrent}
Shaikh Z.~I.,  Raghav A.,  Vichare G.,  Bhaskar A.,  Mishra W.,   Choraghe K.,
  2019, MNRAS, 490, 3440

\bibitem[\protect\citeauthoryear{Tatrallyay, Russell, Luhmann, Barnes  \&
  Mihalov}{Tatrallyay et~al.}{1984}]{Tatrallyay1984}
Tatrallyay M.,  Russell C.~T.,  Luhmann J.~G.,  Barnes A.,   Mihalov J.~D.,
  1984, \mn@doi [JGR: Space Physics] {10.1029/ja089ia09p07381}, 89, 7381

\bibitem[\protect\citeauthoryear{Totten, Freeman  \& Arya}{Totten
  et~al.}{1995}]{totten1995empirical}
Totten T.,  Freeman J.,   Arya S.,  1995, JGR: Space Physics, 100, 13

\bibitem[\protect\citeauthoryear{Van~Doorsselaere, Wardle, Del~Zanna, Jansari,
  VERwICHTE  \& NAkARIAkOV}{Van~Doorsselaere et~al.}{2011}]{van2011first}
Van~Doorsselaere T.,  Wardle N.,  Del~Zanna G.,  Jansari K.,  VERwICHTE E.,
  NAkARIAkOV V.~M.,  2011, ApJL, 727, L32

\bibitem[\protect\citeauthoryear{Verscharen, Chen  \& Wicks}{Verscharen
  et~al.}{2017}]{Verscharen_2017}
Verscharen D.,  Chen C. H.~K.,   Wicks R.~T.,  2017, The Astrophysical Journal,
  840, 106

\bibitem[\protect\citeauthoryear{Wang, Wang, Shen, Shen  \& Lugaz}{Wang
  et~al.}{2014}]{wang2014deflected}
Wang Y.,  Wang B.,  Shen C.,  Shen F.,   Lugaz N.,  2014, Journal of
  Geophysical Research: Space Physics, 119, 5117

\bibitem[\protect\citeauthoryear{Wang, Ofman, Sun, Provornikova  \&
  Davila}{Wang et~al.}{2015}]{wang2015evidence}
Wang T.,  Ofman L.,  Sun X.,  Provornikova E.,   Davila J.~M.,  2015, ApJL,
  811, L13

\bibitem[\protect\citeauthoryear{Wang et~al.,}{Wang
  et~al.}{2016}]{wang2016propagation}
Wang Y.,  et~al., 2016, Journal of Geophysical Research: Space Physics, 121,
  7423

\bibitem[\protect\citeauthoryear{Winslow et~al.,}{Winslow
  et~al.}{2016}]{winslow2016longitudinal}
Winslow R.~M.,  et~al., 2016, Journal of Geophysical Research: Space Physics,
  121, 6092

\bibitem[\protect\citeauthoryear{Zhu}{Zhu}{1990}]{zhu1990plasma}
Zhu X.,  1990, GRL, 17, 2321

\bibitem[\protect\citeauthoryear{Zhuang, Lugaz, Gou, Ding  \& Wang}{Zhuang
  et~al.}{2020}]{zhuang2020role}
Zhuang B.,  Lugaz N.,  Gou T.,  Ding L.,   Wang Y.,  2020, The Astrophysical
  Journal, 901, 45

\bibitem[\protect\citeauthoryear{Zurbuchen \& Richardson}{Zurbuchen \&
  Richardson}{2006}]{zurbuchen2006situ}
Zurbuchen T.~H.,  Richardson I.~G.,  2006, Coronal mass ejections, pp 31--43

\makeatother
\end{thebibliography}






\end{document}